\documentclass[12pt]{article}
\pdfoutput=1
\usepackage[utf8]{inputenc}
\usepackage{amsmath, amsthm, amsfonts}
\usepackage{graphicx}
\usepackage{natbib}
\usepackage{url}
\usepackage[english]{babel}
\usepackage[T1]{fontenc}
\usepackage{fancybox}
\usepackage{lmodern}
\usepackage{multicol}
\usepackage{cancel}
\usepackage{bbm}
\usepackage{float}
\usepackage[ruled,vlined]{algorithm2e}

\theoremstyle{definition}
\usepackage{multirow}\usepackage{booktabs}

\DeclareMathOperator*{\argmax}{argmax}

\usepackage[top=2.5cm, bottom=2.5cm, left=2.5cm, right=2.5cm]{geometry}

\begin{document}

  \title{\bf Random forests for high-dimensional longitudinal data}
  \author{Louis Capitaine, Robin Genuer and Rodolphe Thi\'{e}baut\thanks{
    The authors gratefully acknowledge the DALIA trial group}\hspace{.2cm}\\
    INSERM U1219 Bordeaux Population Health Research Center,\\
    Bordeaux University, INRIA Bordeaux Sud-Ouest, SISTM team}
  \maketitle

\begin{abstract}
Random forests is a state-of-the-art supervised machine learning method which behaves well in high-dimensional settings although some limitations may happen when $p$, the number of predictors, is much larger than the number of observations $n$. Repeated measurements can help by offering additional information but no approach has been proposed for high-dimensional longitudinal data. Random forests have been adapted to standard (i.e., $n > p$) longitudinal data by using a semi-parametric mixed-effects model, in which the non-parametric part is estimated using random forests. We first propose a stochastic extension of the model which allows the covariance structure to vary over time. Furthermore, we develop a new method which takes intra-individual covariance into consideration to build the forest. Simulations reveal the superiority of our approach compared to existing ones. The method has been applied to an HIV vaccine trial including 17 HIV infected patients with 10 repeated measurements of 20000 gene transcripts and the blood concentration of human immunodeficiency virus RNA at the time of antiretroviral interruption. The approach selected 21 gene transcripts for which the association with HIV viral load was fully relevant and consistent with results observed during primary infection.
\end{abstract}

\vfill

\section{Introduction}
\label{sec:intro}

Random forests (RF henceforth), introduced by \cite{breiman2001random}, is a non-parametric statistical learning method from the ensemble decisions trees approaches which is one of the state-of-the-art machine learning method for prediction and classification \citep{fernandez2014we}.
It shows a good behavior for applications in high-dimensional settings where the number of predictors $p$ is larger than the number of observations $n$ \citep[e.g.,][]{cutler2007random, chen2012random}.

Theoretical background of RF has also been recently improved \citep{scornet2015consistency, mentch2016quantifying, wager2014asymptotic, biau2016random}. \cite{scornet2015consistency} proved a consistency result for RF in the context of additive regression models. \cite{mentch2016quantifying} and \cite{wager2014asymptotic} study asymptotic normality of RF predictions and hence proposed confidence intervals for those predictions. We refer to \cite{biau2016random} for further reading on that matter.

However, some limitations may happen when $p$ is much larger than $n$ \citep{statnikov2008comprehensive}, i.e. $p >> n$. In this situation, parameters, specifically the number of variables randomly picked at each node of a tree, must be carefully tuned to control the randomness of the method \citep{genuer2008random}. Some recent improvements have been suggested to deal specifically with high-dimensional data \citep{zhu2015reinforcement, linero2018bayesian}. \cite{zhu2015reinforcement} used ideas from reinforcement learning  with the tree-based model framework to focus on relevant variables.
\cite{linero2018bayesian} developed Bayesian regression trees
\citep{chipman1998bayesian} where sparsity was implemented by using appropriate priors.

Besides methodological developments to deal with the issue of high dimension, the situation may be improved by additional observations available in each individual from repeated measurements leading to longitudinal data.
There is already a large bulk of work on RF for survival (i.e. censored) data \cite{hothorn2005survival, ishwaran2008random, ishwaran2010highdimensional, steingrimsson2018censoring}. \cite{hothorn2005survival} proposed a general survival ensembles framework based on inverse probability censoring weighted loss function, while \cite{ishwaran2008random, ishwaran2010highdimensional} introduced random survival forests by using a split criterion adapted to censored data, and then derive variable selection strategies for high-dimensional survival data.
More recently, \cite{steingrimsson2018censoring} proposed a new survival trees and forests based on the censoring unbiased transformations theory.

Less has been done to adapt random forest approaches to repeated measurement settings. The analysis of repeated measurements requires to take into account the specific correlation structure as done with mixed effects models \citep{laird1982random, verbeke2009linear}. The first approaches dealing with longitudinal and clustered data involved tree-based methods \citep{segal1992tree,
hajjem2011mixed, sela2012re}.

Then, \cite{hajjem2014mixed} proposed a method based on RF. The main idea was to iterate between the fixed part and the random part to estimate the parameters through an Expectation Maximization (EM) algorithm. All these approaches can be viewed as semi-parametric mixed effects model where the fixed effects part is modeled through tree-based methods. Of note, other approaches based on smoothing splines have been proposed for semi-parametric mixed effects model such as in \cite{jacqmin2002penalized, zhang1998semiparametric, wang1998smoothing}.

The main contributions of this paper are three-fold: first we extend existing methods by using a stochastic semi-parametric mixed effects model (Section~\ref{sec:model}).
Secondly, we introduce a new method based on RF to handle high-dimensional longitudinal data and derive theoretical guarantees for predictions of our model in an asymptotic framework (Section~\ref{sec:estimate}).
Third, we compare existing methods with ours in an extensive simulation study (Section~\ref{sec:simu}) and analyze a therapeutic vaccine trial in HIV-infected patients in Section~\ref{sec:appli}.

All existing and proposed methods have been implemented together in an R package called \texttt{longituRF}\footnote{Available at \url{https://github.com/Lcapitaine/longituRF}}.

\section{The semi-parametric stochastic mixed effects model}
\label{sec:model}

Let us consider longitudinal data with $n$ individuals, the $i$th individual having $n_{i}$ observations over time. Suppose $Y_{ij}$ (for all $i=1,...,n$ and $j=1,...,n_{i}$), the response of the $i$th individual at time $t_{ij}$, satisfies 
\begin{equation}
Y_{ij}=f(X_{ij})+Z_{ij}b_{i}+\omega_{i}(t_{ij})+\varepsilon_{ij}
\label{eq:defSemiParaMod}
\end{equation}
where $X_{ij}$ is the $p\times 1 $ vector of covariates, $f: \mathbb{R}^{p}\longrightarrow\mathbb{R}$
is the unknown mean behavior function, $b_{i}$ is a $q\times 1$ vector of random effects associated with the $1\times q$ vector of  covariates $Z_{ij}$, $\omega_{i}(t)$ is a stochastic process used to model serial correlation and  $\varepsilon_{ij}$ denotes a measurement error. The $b_{i}$, are independent, as well as the $\omega_{i}\left(t\right)$ and the $\varepsilon_{ij}$ for all $i=1,...,n;\ j=1,...,n_{i}$.  We also assume that $b_{i},\omega_{i}(t)$ and $\varepsilon_{ij}$ are mutually independent.

We suppose that the $b_{i}$ are normally distributed as $\mathcal{N}(0,B)$ where $B$ is a $q\times q$ positive definite matrix; $\omega_{i}(t)$ is a centered Gaussian process with covariance function $\Gamma_{i}(s,t;\gamma^{2})=cov\left(\omega_{i}(t),\omega_{i}(s)\right)$ depending on a parameter $\gamma^{2}$; and the $\varepsilon_{ij}$ are normally distributed as $\mathcal{N}(0,\sigma^{2})$. 

We consider in model (1) that the evolution of the response variable for the $i$th individual $Y_{i}$ over time varies around a mean behavior function that is given by $f$. These variations specify the individual trajectories around $f$ and are given by the random effects $b_{i}$ and the stochastic process $\omega_{i}(t)$ for the $i$th individual. 

We suppose here that $p$ the number of covariates for the mean behavior is much larger than $N=\sum_{i=1}^{n}n_{i}$ the total number of observations, this leads us in a high dimensional context. \cite{zhang1998semiparametric} considered a semi-parametric stochastic model close to model (1) in small dimension with $f$ a $1-$dimensional function of the time, hence (1) can be seen as a generalization of their model.
\cite{hajjem2014mixed} considered a model similar to (1) but without the stochastic processes $\omega_{i}(t)$, and developed a method based on the EM algorithm \citep{mcculloch1997maximum} to estimate the mean behavior function $f$ with any regression method (splines, RF, kernel-based methods, etc...). Note that if the function $f$ is assumed linear then model (1) reduces to the linear stochastic model of \cite{diggle1989spline}.

\section{Estimation}
\label{sec:estimate}

We write model~\eqref{eq:defSemiParaMod} in the vectorized form as follows,
for all $i = 1, \ldots, n$:
\begin{equation}
Y_{i}=f_{i}+Z_{i}b_{i}+\omega_{i}+\varepsilon_{i}
\label{eq:defSemiParaModVec}
\end{equation}
where $f_{i}=\left(f\left(X_{i1}\right),...,f\left(X_{in_{i}}\right)\right)^{T}$,
$Y_{i}=(Y_{i1},...,Y_{in_{i}})^{T}$, 
$Z_{i}=\left[Z_{i1},...,Z_{in_{i}}\right]^{T}$, \\
$\omega_{i}=\left(\omega_{i}(t_{i1}),...,\omega(t_{in_{i}})\right)^{T}$
and $\varepsilon_{i}=\left(\varepsilon_{i1},...,\varepsilon_{in_{i}}\right)^{T}$.

We suppose that the covariance matrix $\left(\Gamma_{i}\left(t_{ij},t_{ik};\gamma^{2}\right)\right)_{1\leq j,k\leq n_{i}}$ of the process $\omega_{i}$ verifies $\Gamma_{i}\left(t_{ij},t_{ik};\gamma^{2}\right)=\gamma^{2}K_{i}\left(t_{ij},t_{ik}\right)$ with $K_{i}$ a positive definite matrix only depending on observations time for all $1\leq j,k\leq n_{i}$, $i=1,...,n$. We also consider the case where $K_{i}$ depends on an additional parameter $\alpha$ in section 3.2. 

In the same spirit of \cite{hajjem2014mixed}, we use a variant of the EM algorithm to estimate parameters. The main principle of the method is given in the Algorithm 1, while a detailed version can be found in the supplementary materials. 

\begin{algorithm}
    \textbf{initialization}: Let $r=0$, $\widehat{b}_{i,(r)}=0_q$, $\widehat{\omega}_{i,(r)}=0_{n_{i}}$, $\widehat{B}_{(r)}=I_{q}$, $\widehat{\gamma}_{(r)}^{2}=1$ and $\widehat{\sigma}_{(r)}^{2}=1$ ;

    \Repeat{convergence}{
        \begin{enumerate}
            \item $r=r+1$, estimate $f$ in the standard regression mode: $$\widetilde{Y}_{ij, (r-1)}=f\left(X_{ij}\right)+\varepsilon_{ij}$$ where  $\widetilde{Y}_{ij, (r-1)}=Y_{ij}-Z_{ij}\widehat{b}_{i,(r-1)}-\widehat{\omega}_{ij,(r-1)}$.
Then  for given $\widehat{B}_{(r-1)}$, $\widehat{\gamma}_{(r-1)}^{2}$ and \\
$\widehat{\sigma}_{(r-1)}^{2}$ predict $\widehat{b}_{i,(r)}$ and $\widehat{\omega}_{i,(r)}$ for all $i=1,...,n$\;
            
            \item for given $\widehat{f}$, $\widehat{b}_{i,(r)}$ and $\widehat{\omega}_{i,(r)}$, update $\widehat{B}_{(r)}$, $\widehat{\gamma}_{(r)}^{2}$ and $\widehat{\sigma}_{(r)}^{2}$
        \end{enumerate}
    }
    
    \caption{General method}
\end{algorithm}

\subsection{Mean behavior function estimation}
\label{sec:estmoy}

At step 1 of Algorithm 1, we consider variance parameters known and given by estimations of the previous iteration. The mean behavior function $f$ can be estimated with any regression method. When $f$ is estimated with CART tree,
\cite{hajjem2011mixed} refer to  \textbf{MERT} (\textbf{M}ixed \textbf{E}ffects \textbf{R}andom \textbf{T}rees).
CART \citep{breiman1984classification} consists in partitioning in a recursive way the explanatory variable space to obtain the best partition for prediction. At each step of the partitioning, the space is cut into two sub-parts. Hence, the obtained partition can naturally be associated to a binary tree which is called CART tree. Furthermore, we stress that each split is optimized among all explanatory variables and that the CART algorithm works with two steps: the maximal tree building following by the pruning step, in order to give the best predictor in terms of prediction error.

Similarly \cite{hajjem2014mixed} refer to \textbf{MERF} (\textbf{M}ixed \textbf{E}ffects \textbf{R}andom \textbf{F}orest) when $f$ is estimated with RF. RF are an aggregation of multiple randomized CART trees, where the aggregation consists in tacking the mean of individual trees predictions. Each tree is a maximal tree built using a random perturbation: first, it is built on a bootstrap sample of the learning set, and secondly, at each step of the partitioning, the best split is optimized among a randomly drawn subset of explanatory variables. The size of the subset of variables, often called \texttt{mtry}, is the most important parameter of the method.
RF naturally estimate the prediction error with the Out-Of-Bag (OOB) error as
the following: to predict the response of one particular observation of the
learning set, only trees built on bootstrap samples not containing this
observation are aggregated.
Furthermore, OOB samples (made of observations not selected in bootstrap
samples) are also used to compute a variable importance (VI) score. For a fixed
variable, the VI score of this variable is defined as the mean increase of the
error of a tree on its associated OOB sample after a random permutation of this
variable values.

For the considered model (1) we denote by \textbf{SMERF} (\textbf{S}tochastic \textbf{M}ixed \textbf{E}ffects \textbf{R}andom \textbf{F}orests) the generalization of \textbf{MERF} that we propose.

When the mean behavior function is estimated with a CART tree $T$, \cite{sela2012re} proposed to use the partition associated with $T$ to fit a linear mixed effects model.
Let $\Phi^{i}$ the indicator matrix defined as $\Phi^{i}_{jk}=\mathbbm{1}_{\{X_{ij}\in g_{k}\}}$ where $g_{k}$ is the $k$th leaf of the tree $T$. Considering the following model
$$Y_{i}=\Phi^{i}\mu_{T}+Z_{i}b_{i}+\omega_{i}+\varepsilon_{i} \enspace ,$$
the estimation of the values $\mu_{T}$ of the associated leafs of $T$ at step 1 are given by
$$\widehat{\mu}_{T}=\left(\underset{1\leq i\leq N}\sum(\Phi^{i})^{T}V_{i}^{-1}\Phi^{i}\right)^{-1}\left(\underset{1\leq i\leq N}\sum(\Phi^{i})^{T}V_{i}^{-1}Y_{i}\right)$$
with $V_{i}=Var(Y_{i})=Z_{i}BZ_{i}^{T}+\gamma^{2}K_{i}+\sigma^{2}I_{n_{i}}$ for all $i=1,...,N$.

With this method, the values associated with the leafs of $T$ are computed by taking in account intra-individual covariance matrix $V_{i}$ instead of taking the simple mean of values in the leaf. The fitted tree $\widehat{f}_{i}=\Phi^{i}\widehat{\mu}_{T}$ is called \textbf{REEMtree}. 

Following this work, we propose a novel method, named \textbf{REEMforest}, which aggregate a collection of \textbf{REEMtree}.
Let $K$ randomized trees $T_1,...,T_K$, $\Phi^{i,k}$ the indicator matrix associated with the $k$th random tree $T_k$ and $\widehat{\mu}_{T_{k}}$ the fitted leafs of $T_k$ estimated with the stochastic linear mixed-effects model $Y_i=\Phi^{i,k}\mu_{T_k}+Z_i b_i+\omega_i+\varepsilon_i$. The \textbf{REEMforest} estimator is given by the mean of the $K$ fitted \textbf{REEMtree}:
 $$\widehat{f}_{i}=\frac{1}{K}\sum_{k=1}^{K}\Phi^{i,k}\widehat{\mu}_{T_{k}} \enspace .$$ 
 Finally, when the considered model is (1) we denote by \textbf{SREEMforest} the method with the additional estimation of the stochastic process.

Once $\widehat{f}_{i}$ has been computed, the predictions for the random effects $b_{i}$ and the stochastic processes $\omega_{i}$ for known parameters $\left(B,\gamma^{2}, \sigma^{2}\right)$ are obtained by taking their conditional expectations given the data $Y_{i}$, the best linear unbiased predictors \textbf{BLUP} are:
\begin{align*}
\widehat{b}_{i}&=BZ_{i}^{T}V_{i}^{-1}\left(Y_{i}-\widehat{f}_{i}\right) \\
\widehat{\omega}_{i}&=\gamma^{2}K_{i}V_{i}^{-1}\left(Y_{i}-\widehat{f}_{i}\right)
\end{align*}

\subsection{Variance components estimation}
\label{sec:varest}

At step 2 of Algorithm 1, the estimation of the variance parameters are obtained by taking the conditional expectation of their maximum likelihood estimators given the data $Y_{i}$.
Thanks to the conditional independence between the individuals we can write,
for fixed $f_i, \, i = 1, \ldots, n$ the likelihood function associated to
model~\eqref{eq:defSemiParaModVec} as follows:
$$\mathcal{L}(B,\gamma^{2},\sigma^{2}; Y) = \prod_{i=1}^{n} \mathcal{L}_i
\left( Y_i;B,\gamma^{2},\sigma^{2} \right) $$
with
$$\mathcal{L}_i \left( Y_i;B,\gamma^{2},\sigma^{2} \right) =
\frac{1}{\left(2\pi\right)^{\frac{n_{i}}{2}}\sqrt{\det(V_{i})}}\exp\{-\frac{1}{2}\left(Y_{i}-f_{i}\right)^{T}V_{i}^{-1}\left(Y_{i}-f_{i}\right)\}$$
the density function on the vector $Y_i$.
Moreover, since $Y_i | b_i,\omega_{i}\sim\mathcal{N}\left(f_{i}+Z_{i}b_{i}
+\omega_{i},\sigma^{2}I_{n_{i}}\right)$, by using the independence of $b_{i}$,
$\omega_{i}$ and $\varepsilon_{i}$ we can easily write the likelihood function
$\mathcal{L}$ as:
\begin{align*}
\mathcal{L}(B,\gamma^{2},\sigma^{2}; Y) &=
\prod_{i=1}^{n} \frac{1}{\left(2\pi\sigma^{2}\right)^{\frac{n_{i}}{2}}}\exp\{-\frac{1}{2\sigma^{2}}\left(Y_{i}-f_i-Z_{i}b_{i}-\omega_{i}\right)\left(Y_{i}-f_i-Z_{i}b_{i}-\omega_{i}\right)^T\}\times \\
& \hspace*{-5ex} \frac{1}{\left(2\pi\right)^{\frac{q}{2}}\sqrt{\det(B)}}\exp\{\frac{1}{2}b_{i}^{T}B^{-1}b_{i}\}\times \frac{1}{\left(2\pi\right)^{\frac{n_{i}}{2}}\sqrt{\det\left(\gamma^{2}K_{i}\right)}}\exp\{\frac{1}{2}\omega_{i}^{T}\left(\gamma^{2}K_{i}\right)^{-1}\omega_{i}\} \enspace .
\end{align*}
Using that $Y_{i}-f_{i}-Z_{i}b_{i}-\omega_{i}=\varepsilon_{i}$, the maximum
likelihood estimators of $B,\gamma^{2}$ and $\sigma^{2}$ are:
\begin{equation*}
\widetilde{B}=\frac{1}{n}\sum_{i=1}^{n}b_{i}b_{i}^{T},
\quad \widetilde{\gamma}^2=\frac{1}{N}\sum_{i=1}^n\omega_{i}^{T}K_{i}^{-1}\omega_{i},
\quad \widetilde{\sigma}^{2}=\frac{1}{N}\sum_{i=1}^{n}\varepsilon_{i}^{T}\varepsilon_{i}
\end{equation*}

Because $b_{i}$, $\omega_{i}$ and $\varepsilon_{i}$ are unknown these estimators are not computable, this is why we take the expectation given the data $Y_{i}$. The conditional expectations of the estimators $\widetilde{B}$ and $\widetilde{\sigma}^{2}$ are given in \cite{wu2006nonparametric}: 
\begin{align*}
\widehat{B} &= \mathbb{E}(\widetilde{B}|Y)=\frac{1}{n}\sum_{i=1}^{n}
\left\{ \widehat{b}_{i}\widehat{b}_{i}^{T}+ B - BZ_{i}^{T}V_{i}^{-1}Z_{i}B\right\} \\
\widehat{\sigma}^{2} &= \mathbb{E}(\widetilde{\sigma}^{2}|Y)
=\frac{1}{N}\sum_{i=1}^{n} \left\{ \widehat{\varepsilon}_{i}^{T}
\widehat{\varepsilon}_{i} + \sigma^{2}tr(V_{i}^{-1}) \right\}
\end{align*}
The conditional expectation of the maximum likelihood estimator $\widetilde{\gamma}^{2}$ given the data $Y_{i}$ is
\begin{align*}
\widehat{\gamma}^{2} &= \mathbb{E}(\widetilde{\gamma}^{2}|Y_{i})=\frac{1}{N}\sum_{i=1}^{n}tr\left(\mathbb{E}(K_{i}^{-1}\omega_{i}\omega_{i}^{T}|Y_{i})\right)\\
&=\frac{1}{N}\sum_{i=1}^{n}tr\left(K_{i}^{-1}\widehat{\omega}_{i}\widehat{\omega}_{i}^{T}+Cov(K_{i}^{-1}\omega_{i},\omega_{i}|Y_{i})\right)\\
&=\frac{1}{N}\sum_{i=1}^{n}\left( \widehat{\omega}_{i}^{T}K_{i}^{-1}\widehat{\omega}_{i}+ \gamma^{2}\left(n_{i}-\gamma^{2}tr\left(V_{i}^{-1}K_{i}\right)\right)\right) \quad \forall i=1,...,n
\end{align*}
Estimators of variance parameters $B$, $\gamma^{2}$ and $\sigma^{2}$ at step 2 are given by $\widehat{B}$, $\widehat{\gamma}^{2}$ and $\widehat{\sigma}^{2}$. 

Gaussian processes such as Ornstein-Uhlenbeck process and fractional Brownian motion have a variance-covariance function $\Gamma_{i}\left(s,t;\gamma^{2},\alpha\right)$ which depends on two parameters $\gamma^{2}$ and $\alpha$. This covariance function $\Gamma_{i}\left(s,t;\gamma^{2},\alpha\right)$ can be written as $\gamma^{2}K_{i}\left(s,t;\alpha\right)$ with $K_{i}$ depending on $\alpha$. There is no analytic maximum likelihood estimator of $\alpha$. However, for a fixed value of $\alpha$, the estimation procedure described in this section holds. Thus for $\mathcal{H}=\{\alpha_{1},...,\alpha_{d}\}$ an ensemble of possible values of $\alpha$ parameter, the estimator of $\alpha$ is 
\begin{equation*}
\widehat{\alpha}=\argmax_{\alpha\in\mathcal{H}} l\left(B,\gamma^{2},\alpha,\sigma^{2};y\right)
\end{equation*}
where $l$ is the log-likelihood function.

\section{Asymptotic analysis}
\label{sec:amp}
\subsection{Convergence of the expected outcome estimation}

Given a fitted \textbf{MERF} forest $\widehat{f}$, the associated random effects $\widehat{b}_{i}$ and the stochastic processes $\widehat{\omega}_{i}$ obtained with the algorithm 1, the fitted response variable $Y_{i}$ for the $i$th individual is 
\begin{equation*}
\widehat{Y}_{i}=\widehat{f}_{i}+Z_{i}\widehat{b}_{i}+\widehat{\omega}_{i}
\end{equation*}
Considering the variance components known, the expectations of random effects $\widehat{b}_{i}$ and $\widehat{\omega}_{i}$ are related to the bias of $\widehat{f}_{i}$ by
\begin{align*}
\mathbb{E}\left(\widehat{b}_{i}\right|B,\gamma^{2},\sigma^{2})&=-BZ_{i}^{T}V_{i}^{-1}Bias\left(\widehat{f}_{i}\right) \\
\mathbb{E}\left(\widehat{\omega}_{i}\right|B,\gamma^{2},\sigma^{2})&=-\gamma^{2}K_{i}V_{i}^{-1}Bias\left(\widehat{f}_{i}\right)
\end{align*}
This leads to 
\begin{equation*}
\mathbb{E}\left(\widehat{Y}_{i}|B,\gamma^{2},\sigma^{2}\right)=f_{i} + \left(I_{n_{i}}-BZ_{i}^{T}V_{i}^{-1}-\gamma^{2}K_{i}V_{i}^{-1}\right)Bias\left(\widehat{f}_{i}\right)
\end{equation*}
Under some regularity conditions on the function $f$, thanks to \cite{scornet2015consistency} we conclude that $\mathbb{E}\left(\widehat{Y}_{i}\right|B,\gamma^{2},\sigma^{2})-f_{i}\underset{n\to+\infty}{\longrightarrow} 0$. 

This means that when the number of individuals is large enough, the mean fit tends to the true mean behavior $f_{i}$.

\subsection{Convergence of out-of-sample outcome predictions}

The prediction of the response variable for the $i$th individual at time $t$ is 
\begin{equation}
\widehat{Y}_{i}(t) = \widehat{f}\left(X_{i}(t)\right)+Z_{i}\left(t\right)\widehat{b}_{i}+\widetilde{\omega}_{i}(t)
\label{eq:predOOS}
\end{equation}
with $X_{i}(t)$ and $Z_{i}\left(t\right)$ the fixed and random effects explanatory variables for the $i$th individual at time $t$ and 
\[\widetilde{\omega}_{i}(t)=
		\left\{
		\begin{array}{ll}
				\frac{1}{t_{+}-t_{-}}\left[(t-t_{-})\widehat{\omega}_{i}(t_{-})+(t-t_{+})\widehat{\omega}_{i}(t_{+})\right]\quad \mbox{if $t_{i,1}\leq t\leq t_{i,n_{i}}$}\\
				\mathbb{E}\left(\omega_{i}(t)|\widehat{\omega}_{i}(t_{i,1})\right)= \frac{K_{i}\left(t,t_{i,1}\right)}{K_{i}\left(t_{i,1},t_{i,1}\right)}\widehat{\omega}_{i}(t_{i,1})\quad \mbox{if $t<t_{i,1}$}\\
				\mathbb{E}\left(\omega_{i}(t)|\widehat{\omega}_{i}(t_{i,n_{i}})\right)= \frac{K_{i}\left(t,t_{i,n_{i}}\right)}{K_{i}\left(t_{i,n_{i}},t_{i,n_{i}}\right)}\widehat{\omega}_{i}(t_{i,n_{i}})\quad \mbox{if $t>t_{i,n_{i}}$}\\
		\end{array}
		\right.
\]
		
with $t_{-}=\max\left(\left\{s\in \{t_{i,1},...,t_{i,n_{i}}\}, s\leq t \right\}\right)$ and $t_{+}=\min\left(\left\{s\in \{t_{i,1},...,t_{i,n_{i}}\}, s\geq t \right\}\right)$

With this definition, if $t_{i,1}<t<t_{i,n_{i}}$ it is easy to check that
\begin{align*}
\mathbb{E} \left( \widetilde{\omega}_i (t) | B, \gamma^2, \sigma^2 \right) = -\frac{\gamma^2}{(t_+ - t_-)} &\Big\{ (t - t_-) K_i (t_-, t_-) V_i^{-1} (t_-) Bias \left( \widehat{f} (X_i (t_-)) \right) + \\
&(t_+ - t) K_i (t_+, t_+) V_i^{-1} (t_+) Bias \left( \widehat{f}(X_i (t_+)) \right) \Big\}
\end{align*}

Similarly, for $t\leq t_{i,1}$ or $t\geq t_{i,n_{i}}$ the expectation of $\widetilde{\omega}_{i}\left(t\right)$ is a linear function of \\
$Bias\left(\widehat{f}\left(X_{i}(t)\right)\right)$. According to \cite{scornet2015consistency}, for a new observation for the individual $i$ at time $t$, 
\begin{equation*}
\mathbb{E}\left(\widehat{Y}_{i}\left(t\right)\right|B,\gamma^{2},\sigma^{2})-f\left(X_{i}(t)\right)\underset{n\rightarrow+\infty}{\longrightarrow}0
\end{equation*}

\section{Simulation study}
\label{sec:simu}

\subsection{Simulation model}

\subsubsection{Explanatory variables}
\label{sec:explVar}

In this section we present the approach used to simulate the data matrix of the explanatory variables $X$.
Our choices are motivated by the characteristics of the data coming from our application, which are transcriptomics data in a phase 1/2 vaccine trial (see Section~\ref{sec:appli} for more details).

Since we are in a high dimensional context, we assume that a large majority of variables are not associated to the response variable $Y$ (also known as a \emph{sparsity} assumption).
In our study, those variables are simulated as i.i.d. random draws from a multivariate Gaussian distribution $\mathcal{N} \left( 0, 3 I_N \right)$, where $I_N$ denotes the identity matrix of size $N$ (recall that $N = \sum_{i=1}^n n_i$ denotes the total number of observations).

Moreover, since we deal with longitudinal data in the context of gene expression, we assume that some explanatory variables vary over time and that some explanatory variables are clustered into groups (which correspond to genes involved in the same biological pathway).
\cite{hejblum2015time} highlighted ten examples of groups of genes with different temporal behaviors in the DALIA trial, and we mimic some of these trends by setting the following six behaviors over time in our simulations:
\begin{equation}
\left\{
\begin{array}{l@{\hskip 5ex}l}
			C_{g_1}(t) = 2.44 - 0.04 \left( t - \frac{3(t-6)^2}{t} \right) &
			C_{g_4}(t) = \cos( \frac{t-1}{3} ) \\
			C_{g_2}(t) = 0.5 t-0.1\left(t-5\right)^{2} &
			C_{g_5}(t) = 0.1 t + \sin( 0.6 t + 1.3 )\\
			C_{g_3}(t) = 0.25 t - 0.05 \left( t - 6 \right)^2 &
			C_{g_6}(t) = -0.1 t^2\\
\end{array}
\right.
\label{eq:tempoBehav}
\end{equation}
The explanatory variables with a temporal behavior are then simulated as follows:
$$ X^{(k)}(t)=C_{g_{(k)}}(t)+\zeta_{k}+\varepsilon_{t} $$
where $g_{(k)}$ is the group of the $k$th covariate $X^{(k)}$; $\zeta_{k}\sim\mathcal{N}\left(0,1\right)$ corresponds to a random translation at the group level and $\varepsilon_{t}\sim\mathcal{N}\left(0,0.4\right)$ is an additional time-dependent variability.

In the following, we investigate two situations with different values of the
total number of variables, $p$, as well as different sizes of each group of
variables with temporal behavior.

\subsubsection{Outcome variable}

The two following models, which are special cases of
model~\ref{eq:defSemiParaModVec}, are used to simulate the outcome variable $Y$.
For all $i = 1, \ldots, n$:
\begin{align}
Y_{i} &= f_i+ b_{0i} + z_i b_{1i} + \varepsilon_i
\qquad \mbox{\textit{(non-stochastic model)}} \label{eq:simuModNonSto} \\
Y_{i} &= f_i + b_{0i} + z_i b_{1i} + \omega_i + \varepsilon_i \qquad
\mbox{\textit{(stochastic model)}} \label{eq:simuModSto}
\end{align}
where $(b_{0i}, b_{1i})^{T} \underset{i.i.d}{\sim} \mathcal{N}\left(0,B\right)$
with $B = \begin{pmatrix} 0.5 & 0.6 \\ 0.6 & 3 \end{pmatrix}$, $\omega_i$ is a
Brownian motion with volatility $\gamma^2 = 0.8$ and $\varepsilon_i
\underset{i.i.d}{\sim} \mathcal{N} (0,\sigma^{2}I_{n_i})$ with
$\sigma^{2} = 0.5$. In these models, the random effects $b_{1}$ is associated
with an exogenous variable $Z = (z_{1}, \ldots, z_{n})^T$, where
$z_i \underset{i.i.d}{\sim} \mathcal{U}([0,3])$ for $i = 1, \ldots, n$.

The mean behavior function depends on the dimension of the simulated data:
\begin{itemize}
    \item In the \emph{low-dimensional case} (with $p = 6$):
    \begin{equation}
    f(x) = 1.3 \times \left(x^{(1)}\right)^{2} + 2\times \left| x^{(2)} \right|^{1/2}
    \label{eq:fLowDim}
    \end{equation}
    
    \item In the \emph{high-dimensional case} (with $p = 8000$ and with at
    least 20 variables in the first two groups of explanatory variables):
    \begin{equation}    
    f(x) = 1.3 \times \left( \frac{1}{20}
            \sum_{g\in g_{1}^{20}} X^{(g)} \right)^2
            + 2 \times \left| \frac{1}{20} \sum_{g\in g_{2}^{20}} X^{(g)}
            \right|^{\frac{1}{2}}
    \label{eq:fHighDim}
    \end{equation}
where $g_{1}^{20}$ and $g_{2}^{20}$ represent two sets of 20 genes randomly picked from the group $g_1$ and $g_2$ respectively.
\end{itemize}
The mean behavior function is actually quite the same in the two situations.
The difference lies in the fact that in the high-dimension case, 40 variables
are related to the response variable, against 2 in the low-dimension case.
It is indeed reasonable, in high-dimension, to assume that several genes coming
from the same group are linked to the mean behavior function $f$.

\subsection{Squared bias and prediction error}

The different methods are compared in terms of squared bias (associated to each
estimated quantity) and prediction performance, computed among $M$ repetitions 
of the simulation.

Squared biases are defined as follows:
\begin{align*}
bias^2\left(\widehat{f}^M\right) &= \frac{1}{n\#\mathbbm{T}}\sum_{t\in\mathbbm{T}}\sum_{i=1}^{n}\left\{\widehat{f}^M(X_i(t))-f(X_i(t))\right\}^2 \\
bias^2(\widehat{B}^M) &= \frac{1}{q^{2}}\underset{1\leq k,l\leq q}\sum\left(\widehat{B}_{kl}^M-B_{kl}\right)^2 \\
bias^2\left(\widehat{\gamma}_M^2\right) &= \left(\widehat{\gamma}_M^2-\gamma^2\right)^2 \qquad
bias^2\left(\widehat{\sigma}_M^2\right) = \left(\widehat{\sigma}_M^2-\sigma^2\right)^2
\end{align*}
with
\begin{itemize}
\item $\mathbbm{T}$ a fixed grid of times (different from the times of
measurements),
\item $\widehat{f}^M\left(X_i(t)\right)=\frac{1}{M}\sum_{m=1}^{M}\widehat{f}^m\left(X_i(t)\right)\quad \forall t\in\mathbbm{T},\forall i=1,...,n$
\item $\widehat{f}^m$ the fitted random forest returned by the algorithm after
convergence, associated with the $m$-th repetition.
\item $\widehat{B}^M=\frac{1}{M}\sum_{m=1}^M\widehat{B}^m,
\quad  \widehat{\gamma}^2_M=\frac{1}{M}\sum_{m=1}^M \widehat{\gamma}^2_m,
\quad \widehat{\sigma}^2_M=\frac{1}{M}\sum_{m=1}^M \widehat{\sigma}^2_m$
\item $\widehat{B}^m$ the estimation of $B$ on the $m$-th repetition and similarly for $\widehat{\gamma}^2_M$ and $\sigma^2_M$.
\end{itemize}

To evaluate prediction performance, data associated to one simulation are
randomly split $M'$ times into a learning set and a test set.
Each test set is obtained by randomly picking two measurements of each
individual.
With $\mathcal{T}_i^\ell$ denoting the index of the $i$-th individual
measurements in the $\ell$-th test set, we define the prediction error as:
$$\frac{1}{2nM'} \sum_{\ell=1}^{M'} \sum_{i=1}^n \sum_{j \in \mathcal{T}_i^\ell}
\left(Y_{ij} - \widehat{Y}_{ij}^\ell \right)^2$$
where $\widehat{Y}_{ij}^\ell$ is the predicted response variable, defined in
\eqref{eq:predOOS}, of the $j$-th measure of the $i$-th individual, given
by the fitted random forest returned by the algorithm after convergence.

\subsection{Results}

The number of individuals $n$, is fixed to $17$ (the same as in the DALIA trial) all along the simulation study, and the number of measurements $n_i$ for the $i$th individual is randomly chosen (with uniform distribution) between $8$ and $11$ for every $i = 1, \ldots, n$, leading to an unbalanced design.
We recall that the total number of observations is denoted by $N = \sum_{i=1}^n n_i$.

\subsubsection{A low-dimensional case}

We start by considering a low-dimensional example where $p = 6$.
Hence, we have 6 explanatory variables in the dataset and all of them
have a temporal behavior (respectively given by
Equation~\eqref{eq:tempoBehav}).
This framework allows to compare different tree-based methods as well as a
linear mixed model for longitudinal data in a standard framework.
First, we simulate one dataset under model~\eqref{eq:simuModNonSto} using the
mean behavior function  $f$ defined in Equation~\eqref{eq:fLowDim} and study
the behavior of the \textbf{MERF} method on that dataset.

\begin{figure}[!ht]
\centering
\includegraphics[width=0.7\textwidth]{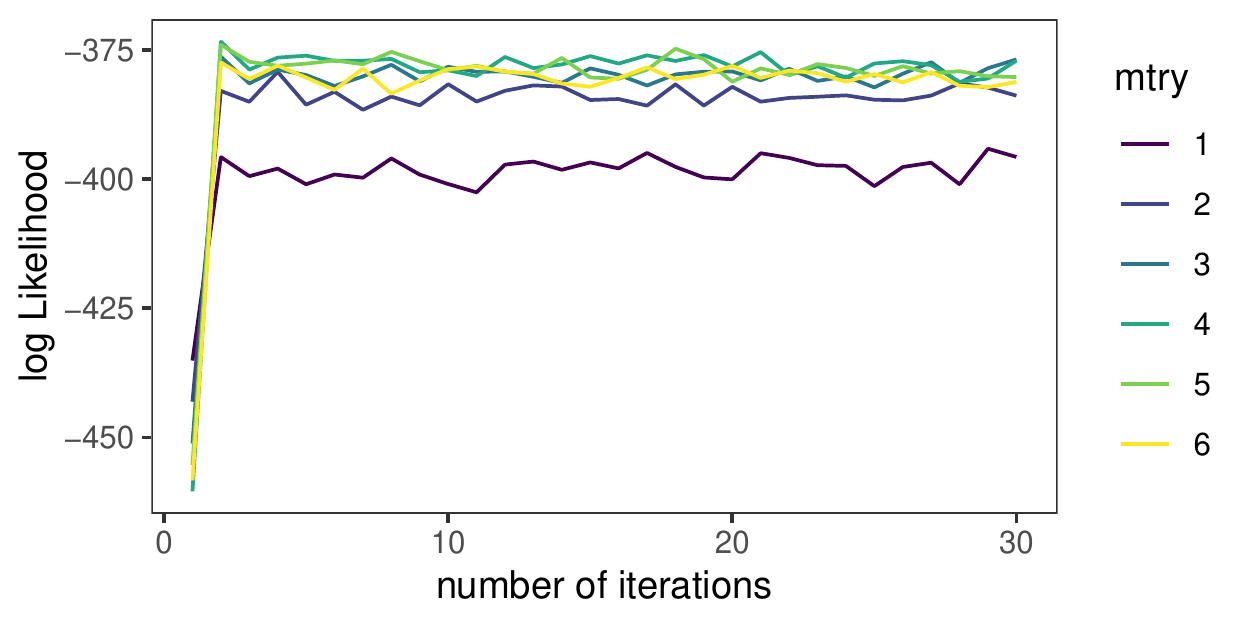}
\caption{Evolution of the log-likelihood against the number of iterations in
\textbf{MERF} method for different \texttt{mtry} values, data simulated under
model~\eqref{eq:simuModNonSto} in the low-dimensional case.}
\label{fig:loglLowDim}
\end{figure}

Figure~\ref{fig:loglLowDim} shows that the convergence of the EM
algorithm for the \textbf{MERF} method is quite affected by the randomness
aspect of the random forests. 
Indeed, the \textbf{MERF} method converge to the maximum likelihood only for
high values of the \texttt{mtry} parameter (the number of variables randomly
drawn before optimizing the split of a node of a tree composing the RF).
It is expected that in an iterative algorithm such as the EM algorithm, an update
with too much randomness compromises the convergence of the algorithm.
Hence, we advocate for large values of \texttt{mtry} in those EM-based RF
methods, even if the trees in the RF are then more similar with each other.

We now simulate 100 datasets (again under model~\eqref{eq:simuModNonSto} with
mean behavior function~\eqref{eq:fLowDim}) and study squared biases on 
estimations of quantities of interest ($f$, $B$, $\sigma^2$ and $\gamma^2$ when
appropriate), given by the four tree-based methods (\textbf{MERT} and \textbf{REEMtree} for 
trees, \textbf{MERF} and \textbf{REEMforest} for forests) and also compare with the Linear Mixed
Effects Model (\textbf{LMEM}) method.
In addition, we simulate 100 additional datasets under the stochastic
model~\eqref{eq:simuModSto} (still with $f$ defined by
Equation~\eqref{eq:fLowDim})
and compare stochastic counterparts of the five methods described above,
respectively denoted by: \textbf{SMERT}, \textbf{SREEMtree}, \textbf{SMERF}, \textbf{SREEMforest} and \textbf{SLMEM}.

\begin{table}[!ht]
\caption{Squared bias of the estimated parameters, averaged on 100 datasets
respectively simulated under model~\eqref{eq:simuModNonSto} and
\eqref{eq:simuModSto} in the low-dimensional case.}
\label{tab:sqBiasLowDim}

\begin{center}
\begin{tabular}{@{}rrrrrc@{}}
\toprule
& $f$ & $B$ & $\gamma^2$ & $\sigma^2$  \\
\midrule
\emph{Non-stochastic model}\\
\textbf{LMEM} & 1.738 & 0.441 & * &  2.517   \\
\textbf{MERT} & 1.831 & 0.254 & * & 0.240    \\
\textbf{REEMtree} & 1.751 & 0.243 & *  & 0.231     \\
\textbf{MERF} & 0.480 & 0.204 & * & 0.009  \\
\textbf{REEMforest} & 0.360 & 0.185 & * & 0.008  \\
\midrule
\emph{Stochastic model}\\
\textbf{SLMEM} & 2.413 & 0.567 & 0.082 & 3.920   \\
\textbf{SMERT} & 2.600 & 0.371 & 0.059 & 1.081   \\
\textbf{SREEMtree}  & 2.969  &  0.321 & 0.001  & 0.361  \\
\textbf{SMERF} & 0.891  & 0.287 &  0.011  & 0.003     \\
\textbf{SREEMforest} & 0.853 & 0.253 & 0.021  & 0.0002  \\
\bottomrule
\end{tabular}
\end{center}

\end{table}

\newpage

As shown in Table~\ref{tab:sqBiasLowDim}, \textbf{LMEM}, \textbf{MERT} and \textbf{REEMtree} methods are
close to each other in terms of bias on $f$ while \textbf{MERF} and \textbf{REEMforest} which use
random forests, provide a much better mean behavior estimation.
Moreover, the squared bias on $f$ for \textbf{REEMforest} is about 25\% lower than
\textbf{MERF} whereas the squared bias on $f$ of \textbf{REEMtree} is only 4\% lower than the one
obtained with \textbf{MERT}.
Hence, in this framework, taking into account the intra-individual covariance structure to
evaluate the tree leafs values generates a much more important decrease of the
squared bias on $f$ when random forests are used compared with CART.
Furthermore, the squared bias obtained on the random effects covariance matrix
$B$ and the residual variance parameter $\sigma^2$ are lower for all four
tree-based methods compared to \textbf{LMEM}, with forests estimating $\sigma^2$ much
better than trees. Finally \textbf{REEMforest} gives slightly lower bias than \textbf{MERF}.

For stochastic methods, tree methods (\textbf{SMERT} and \textbf{SREEMtree}) led to a higher
squared bias of $f$ and lower squared bias of the estimated parameters compared
to \textbf{SLMEM}.
However, forests methods (\textbf{SMERF} and \textbf{SREEMforest}s) still perform better than
trees and \textbf{SLMEM} with squared biases almost all much lower.

\begin{figure}[!ht]
\centering
\includegraphics[width=\textwidth]{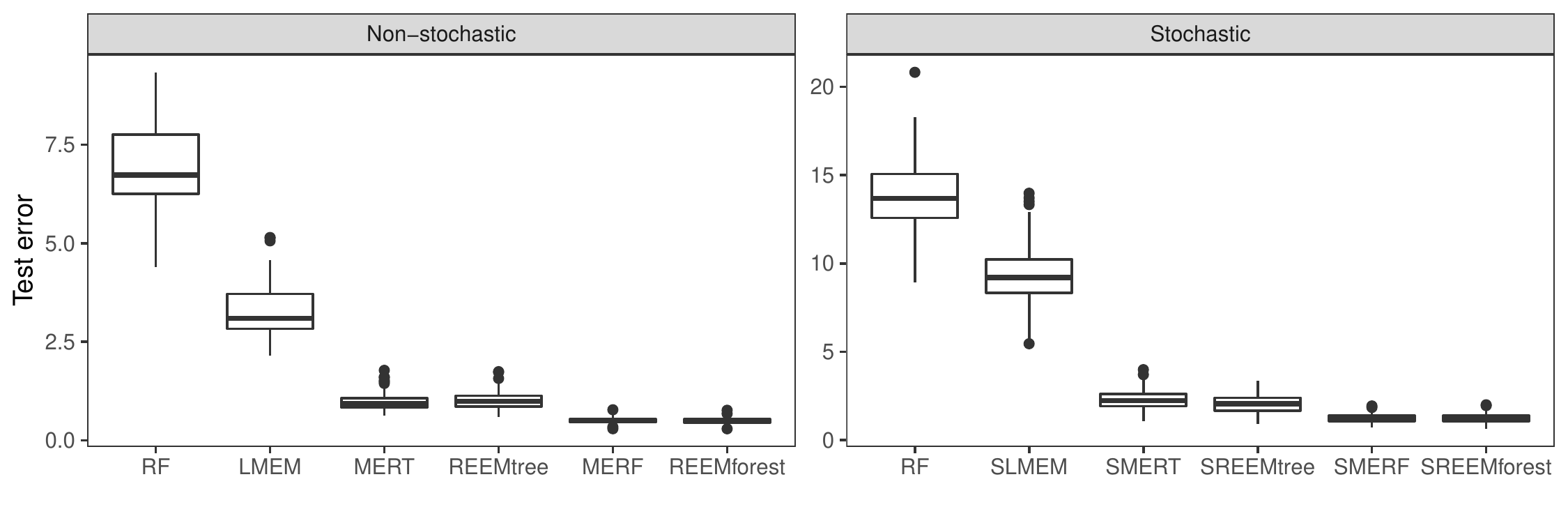}
\vspace*{-5ex}
\caption{Boxplots of the test errors computed on 100 simulated datasets,
either under model~\eqref{eq:simuModNonSto} on the left or
model~\eqref{eq:simuModSto} on the right, in the low-dimensional case.}
\label{fig:boxLowDim}
\end{figure}

Finally, we compare the different methods on their prediction capacity by
computing prediction errors on 100 simulated datasets, either under
model~\eqref{eq:simuModNonSto} or \eqref{eq:simuModSto}.
For each dataset, a test set if formed by randomly picking, for each
individual $i$, two observations among its $n_i$ observations. This gives
a test set containing $2n$ observations and a learning set with $N - 2n$
observations. Breiman's RF is also included in this study in addition to the five methods already mentioned
to illustrate the gain of tacking into account the intra-individual correlation.

As expected, left part of Figure~\ref{fig:boxLowDim} shows that Breiman's RF
performed the worse in terms of prediction error, because it assumes that all
observations are independent, and that \textbf{LMEM} reached a poor prediction
ability, because $f$ is not linear. \textbf{MERT} and \textbf{REEMtree} gave intermediate performance, whereas \textbf{MERF} and
\textbf{REEMforest} reached the lowest prediction errors, with similar performances.
Overall, those comments remains valid for the stochastic case, illustrated
on the right part of Figure~\ref{fig:boxLowDim}.

As a conclusion, we demonstrate the benefits of RF approaches for longitudinal
data analysis in a low-dimensional case, especially in terms of prediction error.
\textbf{REEMforest} exhibited a slight advantage compared to
\textbf{MERF} in terms of validity of the estimation of the mean behavior function
$f$ and of the other parameters $B$, $\sigma^2$ and $\gamma^2$.

\subsection{A high-dimensional case}

For the high-dimensional context, we kept $n = 17$ but set
$p=8000$. We also set the size of each of the six groups containing explanatory
variables with temporal behaviors (given by Equation~\ref{eq:tempoBehav}) to
266, leading to a total of 1596 variables that changed over time among the 8000
variables in the dataset.

First of all, according to Figure~\ref{fig:loglLowDim} for the low-dimensional
case and some preliminary experiments, we fix the \texttt{mtry} parameter of RF
to $5000$ in all RF runs. This ensures convergence of EM-based methods and
avoid a too heavy computational burden compared with a systematic optimization
of \texttt{mtry} on several values for each RF.

\begin{table}[!ht]
\caption{Squared bias of the estimated parameters, averaged on 50 datasets
respectively simulated under model~\eqref{eq:simuModNonSto} and
\eqref{eq:simuModSto} in the high-dimensional case.}
\label{tab:sqBiasHighDim}

\begin{center}
\begin{tabular}{@{}rrrrr@{}}
\toprule
& $f$ & $B$ & $\gamma^2$ & $\sigma^2$  \\
\midrule
\textit{Non-stochastic scheme}\\
\textbf{MERT} & 2.967 & 0.359 & * & 0.596  \\
\textbf{REEMtree} & 2.658 & 0.292 & *  &  0.475\\
\textbf{MERF} & 0.941 & 0.269 & * &  0.216\\
\textbf{REEMforest} & 0.917 & 0.260  & * &  0.211 \\
\midrule
\textit{Stochastic scheme}\\
\textbf{SMERT} & 7.191 & 0.807 & 0.019 & 1.767 \\
\textbf{SREEMtree} & 4.861  & 0.596 & 0.007 & 0.487  \\
\textbf{SMERF} & 1.768 & 0.353 & 0.00001  & 0.562 \\
\textbf{SREEMforest} & 1.754 & 0.333 &  0.0001 & 0.549 \\
\bottomrule
\end{tabular}
\end{center}

\end{table}

We simulated 50 datatsets under model~\eqref{eq:simuModNonSto} (and 50 other
dataset under model~\eqref{eq:simuModSto}) with the mean behavior function
given by Equation~\eqref{eq:fHighDim}, and computed squared biases on
estimations given by the four tree-based methods: \textbf{(S)MERT},
\textbf{(S)REEMtree}, \textbf{(S)MERF} and \textbf{(S)REEMforest}.
We did not compare anymore with \textbf{LMEM} which is not adapted to the
high-dimensional setting.

\begin{figure}[!ht]
\centering
\includegraphics[width=\textwidth]{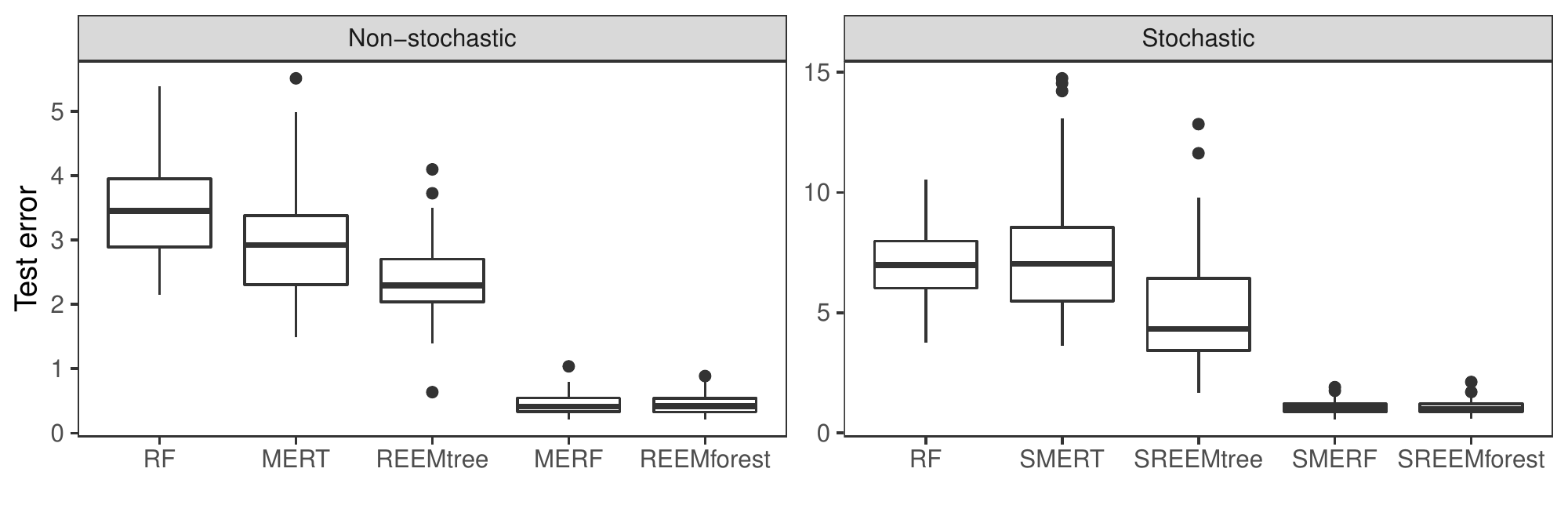}
\vspace*{-5ex}
\caption{Boxplots of test errors computed on 50 simulated datasets,
either under model~\eqref{eq:simuModNonSto} on the left or
model~\eqref{eq:simuModSto} on the right, in the high-dimensional case.}
\label{fig:boxHighDim}
\end{figure}

For the non-stochastic scheme, the squared bias on $f$ and all parameters 
obtained with \textbf{REEMforest} were slightly lower than the one obtained 
with the existing \textbf{MERF} method.
For the stochastic model (18), the two forest-based methods gave similar bias on all parameters.
As in the low dimensional context, forests led systematically to lower
biases on all estimations compared to trees, especially for the estimation of $f$. However in this high-dimensional setting,
\textbf{(S)MERF} and \textbf{(S)REEMforest} performed approximately the same.

We estimated prediction errors of the various methods, by randomly 
selecting test sets in each of the simulated datasets (in the same manner as in
the low-dimensional case in the previous section).
As illustrated in Figure~\ref{fig:boxHighDim}, forests reached very low
prediction error estimations compared to trees.
This last result was expected because it is
well-known that RF performs better than trees for high-dimensional data.
It can also be seen that Breiman's RF (which assume independence between all
observations in the data) are competitive compared with trees, especially
compared with \textbf{(S)MERT}. Hence, in that case, the gain of using RF
instead of trees roughly compensates the fact that Breiman's RF do not
take into account the longitudinal feature of the data.

\begin{figure}[!ht]
\centering
\includegraphics[width=0.5\textheight]{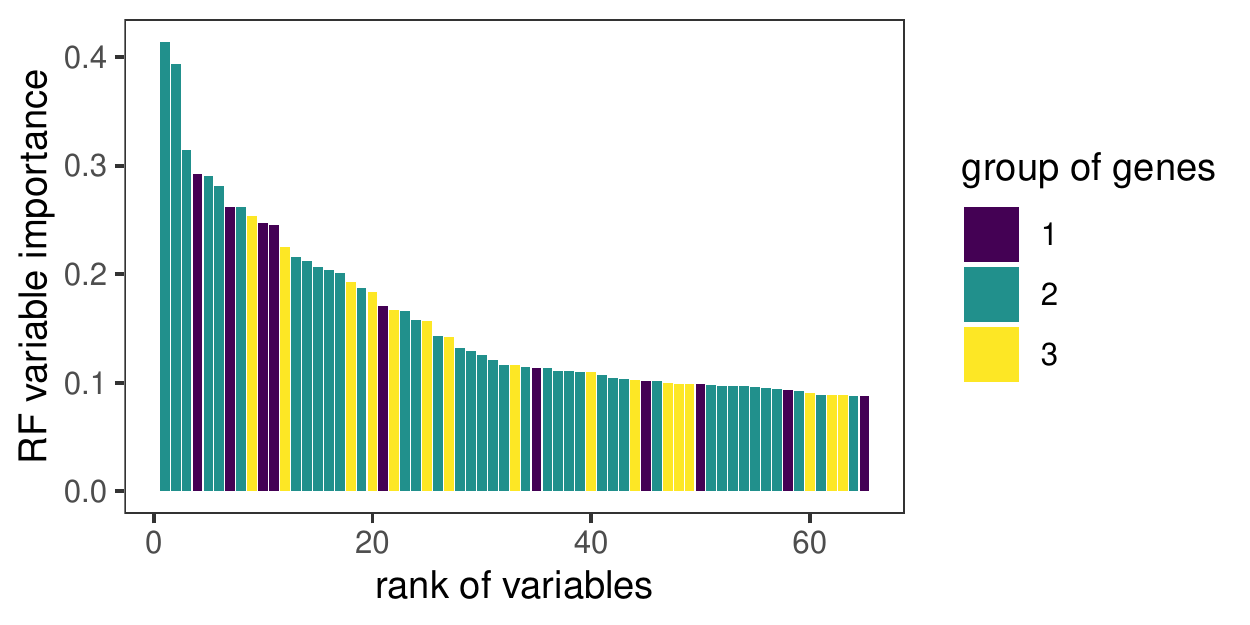}
\caption{Barplot of the first 65 sorted (in decreasing order) variable
importance scores, computed after convergence of the \textbf{REEMforest} method
applied on one dataset simulated under model~\eqref{eq:simuModNonSto}
in the high-dimensional case.}
\label{fig:impREEMfSimu}
\end{figure}

Finally, variable importance (VI) scores computed with the RF returned after convergence
of the \textbf{REEMforest} method are plotted in decreasing order of VI
in Figure~\ref{fig:impREEMfSimu} (only the 65 most important variables are
plotted for the sake of clarity). This graph shows that the most important
variables belong to one of the first three groups of explanatory variables. This result is satisfactory because the mean behavior function (defined by Equation~\ref{eq:fHighDim}) depends on variables that belongs to the first two groups only and the third group is very close to the second one in terms of dynamics (see Equations~\ref{eq:tempoBehav}).

\section{Application to the DALIA vaccine trial}
\label{sec:appli}

DALIA is a therapeutic phase 1/2 vaccine trial including 19 HIV-infected patients who received an HIV vaccine before stopping their antiretroviral treatment (HAART). For a full description of the DALIA vaccine trial we refer to \cite{levy2014dendritic}.

\begin{figure}[!ht]
\centering
\includegraphics[width=0.7\textwidth]{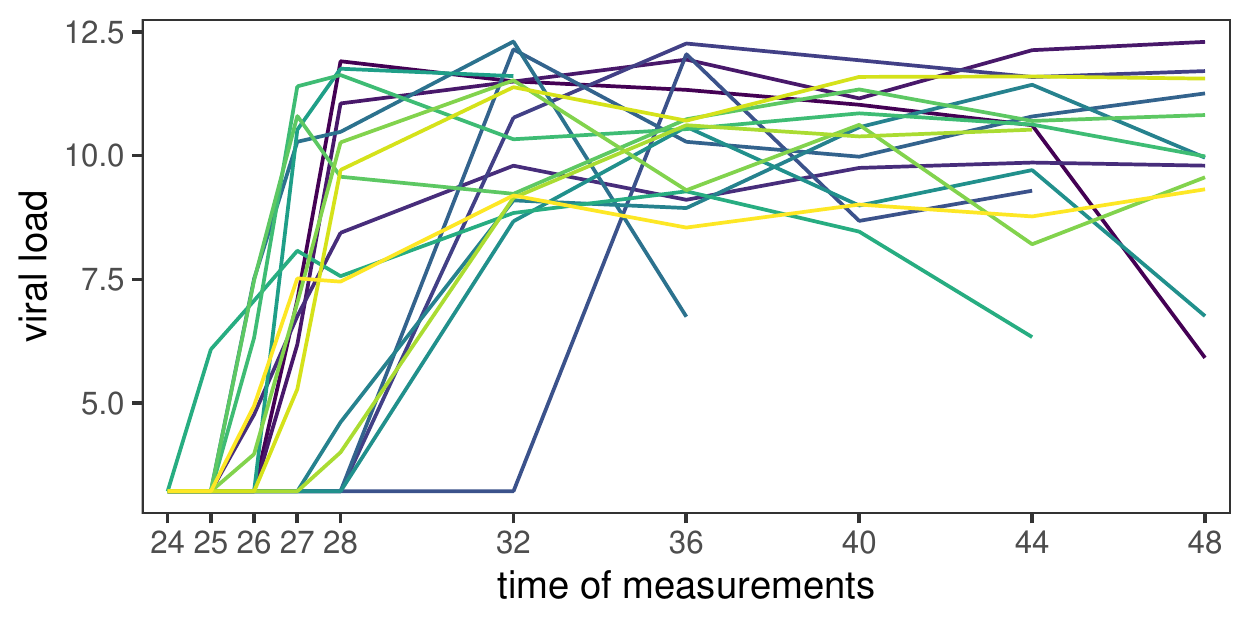}
\caption{Dynamics of plasma HIV viral load (one curve per patient) after antiretroviral treatment interruption, DALIA vaccine trial.}
\label{fig:viralLoad}
\end{figure}

At each harvest time, 32979 gene transcripts were measured as well as the plasma HIV RNA concentration (which was log-transformed) for every patient. In this application, we were interested in finding the gene transcripts associated to the HIV viral load dynamics after antiretroviral treatment interruption. The analysis was performed on the 17 patients with available data at the time of treatment interruption.

Figure~\ref{fig:viralLoad} illustrates the dynamics of the viral replication after antiretroviral treatment interruption with a large between-individuals variability. 

A random intercept and a Gaussian process were included in the model:
\begin{equation}
Y_{ij}=f\left(X_{ij}\right)+b_{0i}+\omega_{i}\left(t_{ij}\right)+\varepsilon_{ij} \quad i = 1, \ldots, n \; ; \; j = 1, \ldots, n_i
\label{eq:modDalia}
\end{equation}
and we will refer to methods only using a random intercept as non-stochastic
methods (\textbf{MERF} and \textbf{REEMforest} in the following).

\begin{figure}[!ht]
\centering
\includegraphics[width=0.7\textwidth]{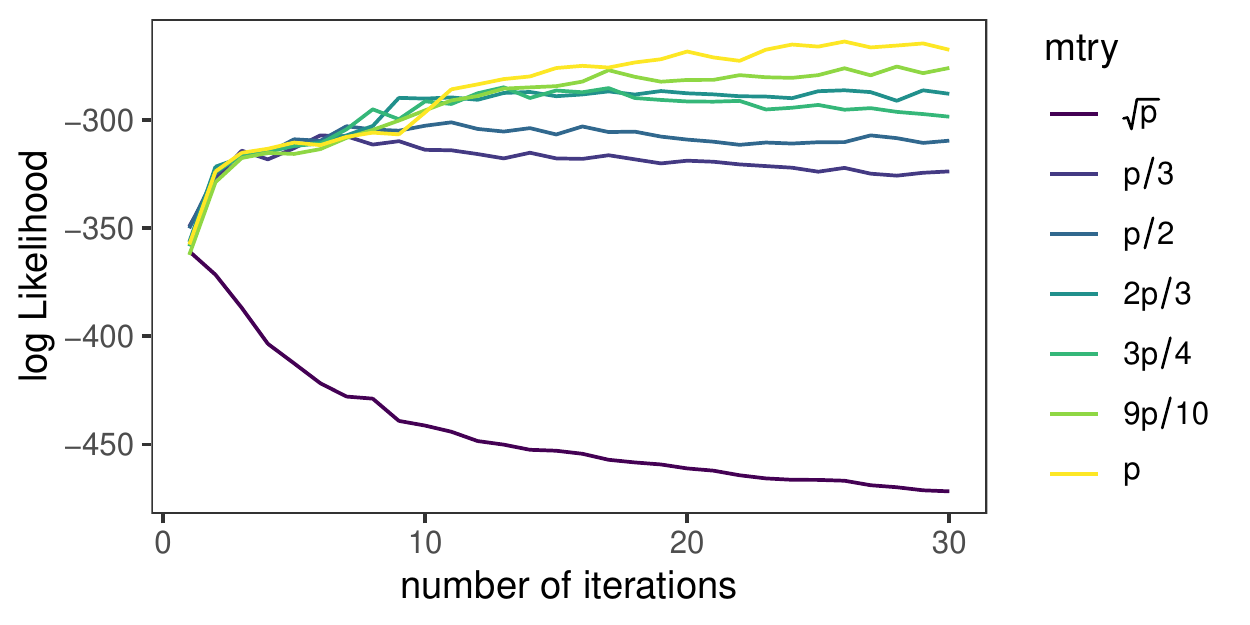}
\caption{Log-likelihood according to the number of iterations in \textbf{SREEMforest} from the model~\eqref{eq:modDalia} with standard Brownian motion, DALIA trial.}
\label{fig:mtryDalia}
\end{figure}

Prediction errors were evaluated with 25 training/test sets random splits. As in the simulation study, a test set was obtained by randomly drawing two observations for each individual.
We chose the stochastic process (between an Ornstein-Uhlenbeck's process and a fractional Brownian motion) that minimized the estimated prediction error.
Hence, the fractional Brownian motion with Hurst exponent $h=\frac{1}{2}$ which is the standard Brownian motion was selected.
Finally, the \texttt{mtry} parameter was fixed to $9p/10 = 29681$ in all experiments of this section, according to the likelihood profile (Figure~\ref{fig:mtryDalia}).

\begin{figure}[!ht]
\centering
\includegraphics[width=0.6\textwidth]{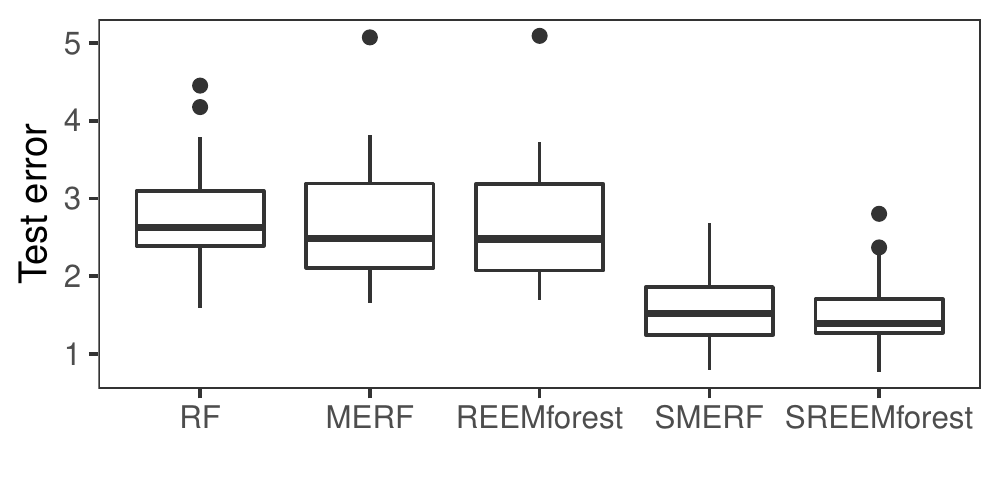}
\vspace*{-2ex}
\caption{Boxplots of test errors computed using 25 training/test sets random
splits, for Breiman's RF, \textbf{MERF}, \textbf{REEMforest}, \textbf{SMERF}
and \textbf{SREEMforest}, DALIA trial.}
\label{fig:boxplotDalia}
\end{figure}

As illustrated in Figure~\ref{fig:boxplotDalia}, Breiman's RF were comparable in terms of prediction error with \textbf{MERF} and \textbf{REEMforest} which only included a random intercept.
However, \textbf{SMERF} and \textbf{SREEMforest} outperformed simple RF and trees, with a slight advantage to \textbf{SREEMforest}.
This confirms, in this real dataset, that taking precisely into account the longitudinal aspect of the data in RF leads to a significant drop of the prediction error.

\subsection{Variable selection using random forests}

Once the algorithm (e.g., \textbf{SREEMforest}) has converged, a variable selection process is applied to select the genes the most associated with the viral load dynamics.
More precisely, the last estimations of $b_{0i}$ and $\omega_i$ (which are outputs of the algorithm) are subtracted from the output variable $Y_i$, for all $i$ (as in step 1 of Algorithm 1) to come back to a classical regression framework (\emph{i.e.}, with independent observations). Hence, the Variable Selection Using Random Forests method from \cite{genuer2010variable} can be apply by using the \texttt{VSURF} package \citep{genuer2015vsurf}.

This method is a fully automatic variable selection procedure based on RF and
designed to deal with high dimensional data in a regression framework as well as
in supervised classification. It works in three steps: i) first, the variables
are sorted in decreasing order of RF variable importance (VI), then a
data-driven threshold is computed to eliminate variables with low VI;
ii) variables left are then introduced (one by one according
to the previous order) 
in nested RF models and the one minimizing the OOB error is selected;
iii) a refined ascending sequence of RF models (obtained in a stepwise way) is
then built and finally the last model of this sequence is returned.

\subsection{Stability of the selected variables set}

We illustrate the stability of the selected variables set by introducing a stability score and studying the behavior of this score against the RF parameter
\texttt{mtry}.

\begin{figure}[!ht]
\centering
\includegraphics[width=0.7\textwidth]{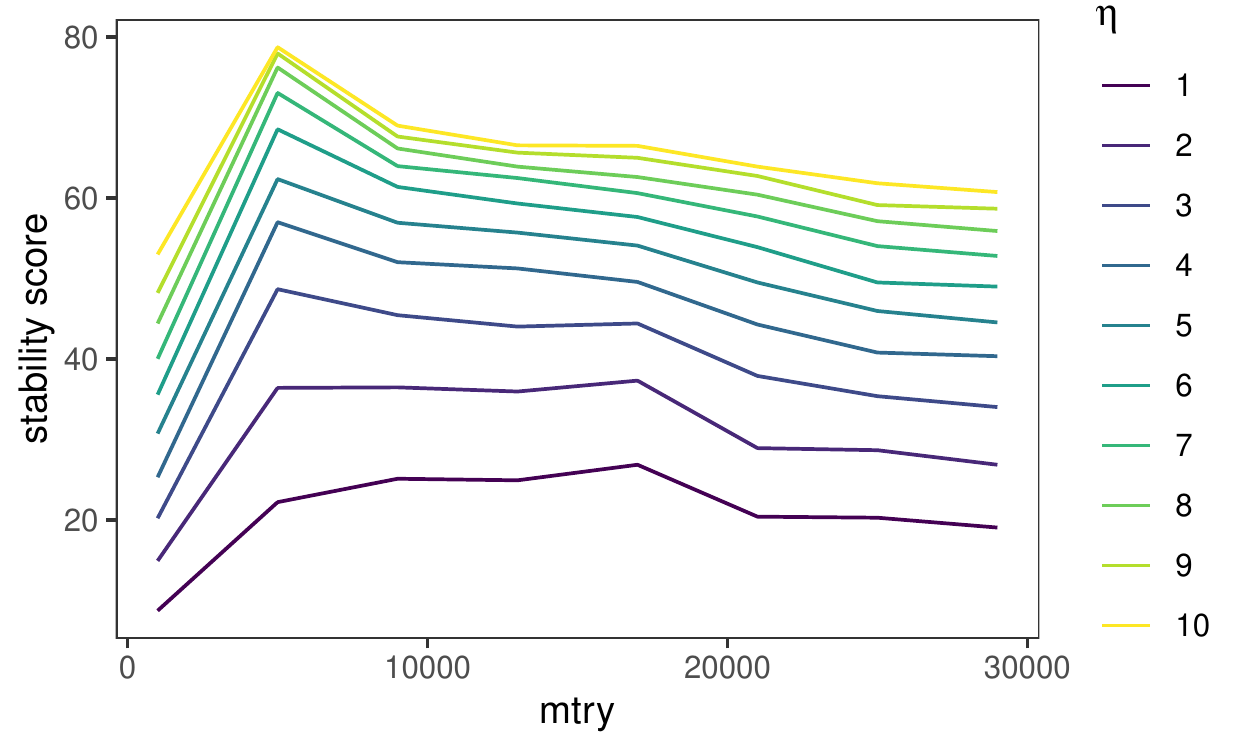}
\caption{Evolution of the mean stability score against the \texttt{mtry} parameter and the neighborhood size ($\eta$), restricted to the 50 most important variables, for the \textbf{SREEMforest} method, DALIA trial.}
\label{fig:stabScores}
\end{figure}

Let $\mathcal{V}= \{V_{(1)},...,V_{(p)} \}$ and $\mathcal{V}'= \{ V_{(1)}',...,V_{(p)}' \}$ be the decreasing ordered variables respectively to the variable importance obtained with two runs of the \textbf{SREEMforest} method.
Due to the randomness aspect of the RF, \textbf{SREEMforest} is random and the sequences $\mathcal{V}$ and $\mathcal{V}'$ may be different.
Hence, we introduce a stability score $\mathcal{SS}$ which measures the difference between two ordered sequence $\mathcal{V}$ and $\mathcal{V}'$:
 $$\mathcal{SS}^{\eta}\left(\mathcal{V},\mathcal{V}'\right)=\frac{1}{p}\sum_{i=1}^p \mathbbm{1}_{\left\{V_{(i)}\in\mathcal{B}\left(V_{(i)}';\eta\right)\right\}}
 \quad \mbox{with} \quad \mathcal{B}\left(V_{(i)}';\eta\right)=\left\{V'_{{(i-\eta)}_+}, \ldots, V'_{{(i+\eta)}_-}\right\}$$
 where $V'_{{(i-\eta)}_+}=
    \begin{cases}
			V_{(1)}' \quad \mbox{if } i-\eta\leq 0\\
			V_{(i-\eta)}'\quad \mbox{else}
    \end{cases}$
and $V'_{{(i+\eta)}_-}=
    \begin{cases}
			V_{(p)} \quad \mbox{if } i+\eta\geq p\\
			V_{(i+\eta)}'\quad \mbox{else}
	\end{cases}
\enspace .$

This score measures the proportion of variables ranked in a same neighborhood ($\eta$ handles the size of the neighborhood). To stabilize the results, the score was computed with 30 pairs of sequences $\mathcal{V}$ and $\mathcal{V}'$ and the mean of the obtained stability scores was provided. 

The computation of these stability scores was restricted to the 50 most important variables given by different runs of \textbf{SREEMforest} applied to the DALIA vaccine trial dataset.
In Figure~\ref{fig:stabScores}, we note that, except for \texttt{mtry} set to 1000, we obtained a stability score around 0.5 for a neighborhood size of 4.
This means that for two lists of the 50 most important variables obtained with \textbf{SREEMforest}, approximately 50\% of them were at the same rank ($\pm 4$
 ranks).
For a neighborhood size larger than 8, the score can exceed 75\%.
In conclusion, for a wide range of \texttt{mtry} values, variable ranking results were quite consistent.

\subsection{Biological results}

The 21 variables selected by \texttt{VSURF} (after convergence of \textbf{SREEMforest}) were mainly transcripts (OAS, LY6E, HERC5, IFI/IFIT, EPSTI1, MX1, RSAD2, EIF2AK2, XAF1) associated to the interferon-$\alpha$ pathway. For instance, they all belongs to the Chaussabel's modules M1.2
and M3.4 annotated "Interferon" \citep{chaussabel2014democratizing}. Interferon pathway is highly correlated to the viral replication as demonstrated previously \citep{bosinger2009global}. Only, two transcripts were not associated to the interferon pathway (EPSTI1 and SAMD9L). The commitment of the interferon pathway reflects the immune response to viral infection. The relevance of these results is another argument for the validation of the proposed approach.

\section{Discussion}

In this article, we introduced a new RF-based methodology suited for the analysis of high-dimensional longitudinal data.
We also generalized existing methods so they can now be used in the stochastic semi-parametric mixed-effects model.
The simulation study revealed the superiority of our approach and of the generalizations we introduced, and the method has been applied to a complex dataset coming from an HIV vaccine trial, illustrating the effectiveness and interest of our approach in such high-dimensional longitudinal context.

One important aspect of the results of our study is the choice of the \texttt{mtry} parameter.
Our advice is to choose a large value for \texttt{mtry}--roughly between $2p/3$ and $3p/4$--, not smaller than $p/2$, and this for two reasons:
first, as we are in an (very) high-dimensional context, the number of variables selected at each node of trees must not be too small--preventing to choose only non-informative variables too often--;
and second, since the different approaches are based on an \emph{iterative} EM-algorithm, the inner RF model must be stabilized--preventing the EM-algorithm to diverge--.
Taking a too small value for \texttt{mtry} could lead to the non-convergence of the method and hence to very sub-optimal results, as illustrated by Figure~\ref{fig:mtryDalia}.

Another key step in these approaches is the choice of the model, and more particularly the choice of the random effects.
Driven by our application, we only use a random intercept (in addition to the stochastic process) regarding the number of individuals and the number of time-points we had in the vaccine trial.
However, in a context with more individuals and/or less time points, it could be interesting to add random effects on the different time points.
This should make the model more flexible and hence increase the capacity of the method to well estimate the inter-individual variability.

There are several issues that require further research.
First, the theoretical analysis of such complex methods (non-parametric estimates plugged into an EM algorithm) seems rather difficult and remains, to the extend of our knowledge, an open issue.
Finally, for the methodological part, another way of adapting RF to longitudinal data could be to change the split criterion of a node in the tree building, as it has been done e.g., for survival RF.

\section*{Appendix}

\begin{algorithm}[H]
\textbf{initialization}: Let $r=0$, $\widehat{b}_{i,(r)}=0$, $\widehat{\omega}_{i,(r)}=0$, $\widehat{B}_{(r)}=I_{q}$, $\widehat{\gamma}_{(r)}^{2}=1$ and $\widehat{\sigma}_{(r)}^{2}=1$ \;

\Repeat{convergence}{
\begin{enumerate}
\item $r=r+1$, for given $\widehat{B}_{(r-1)}$, $\widehat{\gamma}_{(r-1)}^{2}$ and $\widehat{\sigma}_{(r-1)}^{2}$ estimate $f$ on the model $Y_{i}-Z_{i}\widehat{b}_{i,(r-1)}-\widehat{\omega}_{i,(r-1)}$; build $\Phi^{i,k}$ matrices for every tree $T_{k}$ composing \\
the forest and fit the associated leafs $\widehat{\mu}_{T_{k}}$; aggregate trees $$\widehat{f}_{i,(r)}=\frac{1}{K}\sum_{k=1}^{K}\Phi^{i,k}\widehat{\mu}_{T_{k}}$$ then predict $\widehat{b}_{i,(r)}$ and $\widehat{\omega}_{i,(r)}$ $$\widehat{b}_{i,(r)}=\widehat{B}_{(r-1)}Z_{i}^{T}\widehat{V}_{i,(r-1)}^{-1}\left(Y_{i}-\widehat{f}_{i,(r)}\right)$$ $$\widehat{\omega}_{i,(r)}=\widehat{\gamma}_{(r-1)}^{2}K_{i}\widehat{V}_{i,(r-1)}\left(Y_{i}-\widehat{f}_{i,(r)}\right)$$ for all $i=1,...,n$\;

\item for given $\widehat{f}_{i,(r)}$, $\widehat{b}_{i,(r)}$ and $\widehat{\omega}_{i,(r)}$, update $\widehat{B}_{(r)}$, $\widehat{\gamma}_{(r)}^{2}$ and $\widehat{\sigma}_{(r)}^{2}$ $$\widehat{B}_{(r)}=\frac{1}{n}\sum_{i=1}^{n}\{\widehat{b}_{i,(r)}\widehat{b}_{i,(r)}^{T}+\widehat{B}_{(r-1)} - \widehat{B}_{(r)}Z_{i}^{T}\widehat{V}_{i,(r-1)}^{-1}Z_{i}\widehat{B}_{(r)}\}$$ $$\widehat{\gamma}_{(r)}^{2}=\frac{1}{N}\sum_{i=1}^{n}\widehat{\omega}_{i,(r)}^{T}K_{i}^{-1}\widehat{\omega}_{i,(r)}+\widehat{\gamma}_{(r-1)}^{2}\left(n_{i}-\widehat{\gamma}_{(r-1)}^{2}tr\left(\widehat{V}_{i,(r-1)}^{-1}K_{i}\right)\right)$$ $$\widehat{\sigma}_{(r)}^{2}=\frac{1}{N}\sum_{i=1}^{n}\widehat{\varepsilon}_{i,(r)}^{T}\widehat{\varepsilon}_{i,(r)}+\widehat{\sigma}_{(r-1)}^{2}tr\left(\widehat{V}_{i,(r-1)}^{-1}\right)$$ with $\widehat{\varepsilon}_{i,(r)}=Y_{i}-\widehat{f}_{i,(r)}-Z_{i}\widehat{b}_{i,(r)}-\widehat{\omega}_{i,(r)}$ $\forall i=1,...,n$
\end{enumerate}
}
\caption{\textbf{SREEMforest} algorithm}
\end{algorithm}

\newpage

\bibliographystyle{apalike}
\bibliography{RFhigDimLongi_CGT}

\end{document}